\documentclass{lmcs} 


\usepackage{hyperref}
\usepackage{multirow}
\theoremstyle{plain} 

\begin{document}

\title[Strategic analysis of implementation assets and threats]{Strategic analysis of implementation assets and threats}
\titlecomment{{\lsuper*}}

\author[H.~Piech]{Henryk Piech}	
\address{Institute of Computer and Information Sciences, Czestochowa University of Technology, Dabrowskiego 73, 42-201 Czestochowa, Poland}	
\email{henryk.piech@icis.pcz.pl;}  

\author[G.~Grodzki]{Grzegorz Grodzki}	
\address{Institute of Computer and Information Sciences, Czestochowa University of Technology, Dabrowskiego 73, 42-201 Czestochowa, Poland}	
\email{grzegorz.grodzki@icis.pcz.pl;}  
\thanks{}	






\begin{abstract}
  \noindent The aim of the strategic analysis is to (simply) carry out the game \cite{Straffin:mib} between the implementing body and possible links to the existing market situation. We are therefore playing a strategic game between us and the outside world \cite{Owen:mib}. This situation is most often associated with existing and potential threats, such as competition, fashion trends, cost situation, marketing effectiveness, market demand, etc. However, this does not exhaust all aspects resulting from the situation, as conditions conducive to the implementation may also be taken into account, such as the possibility of cooperation, favourable location, new forms of reaching the recipient, new legal solutions, etc. Generally speaking, we can divide the set of conditions into assets and threats. Playing a strategic game leads not only to the estimation of the game's value but also points to the equilibrium points (saddle points)   \cite{Henkelman:mib} and to the conditions of market stabilization \cite{Carfi:mib}. We therefore have two players (one zero-sum two-player game depending on the assumptions made \cite{Fleming:mib}); one side is our assets and the other side is external threats. The strategies of both players will be a combination of implementation, market and marketing parameters. This will be described as an example in the introduction (chapter 1). The next chapters (2 and 3) are proposals for estimating the effects of strategy selection, i.e. creating payoffs for players. The final stage will be playing the game, the analysis of its results (chapters 3) and the summary (chapter 4).
\end{abstract}

\maketitle

\section*{Introduction}\label{S:one}

To describe the strategy, we will use the above mentioned parameters and codify their designation for example as follows: competition (p1), trends (p2), costs (p3), marketing (p4), sales (p5), other (p6).

The values of the parameters are defined in the following ranges: binary, total, real, linguistic, etc. For example, “1” (the binary single) means the strength of an asset and a weak threat. Value “0” (the binary zero) indicates potential weakening leading to a threat. In coding of real numbers we can use percentages or negative fractional numbers in the range of [-1,1], etc. Values of the parameters used in the strategy structure allow to determine the players’ payoffs for particular strategies. For example, for two assets A=\{p1, p2, p3, p4, p5, p6\}=\{1, 0, 1, 1, 1, 0\}, B=\{0, 1, 1, 0, 0, 1\} and three threats C=\{0, 1, 1, 0, 0, 0\}, D=\{1, 0, 1, 0, 1, 1\}, E=\{1, 1, 0, 1, 1, 1\}, we create a game in the form of (Fig. 1):

\begin{figure}[h!]
\centering
{\includegraphics[width=10cm]{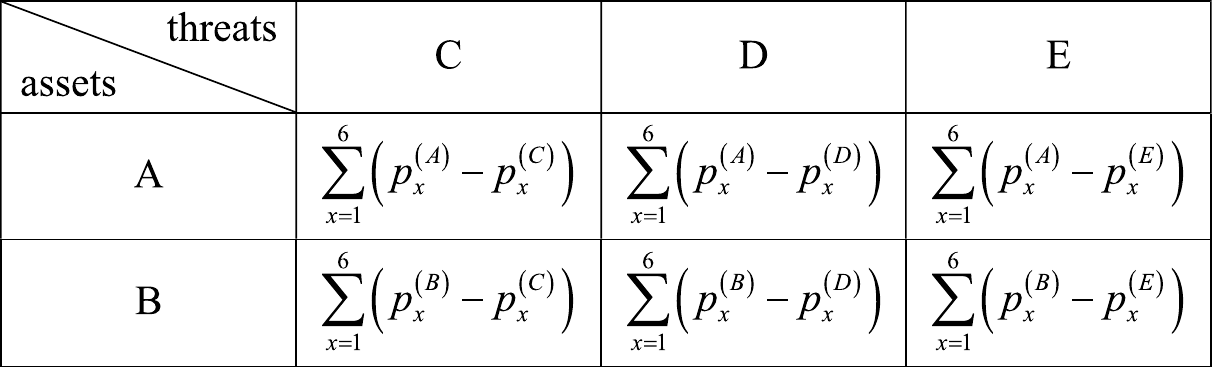}}
\caption{2x3 zero-sum game – formal notation}
\label{fig:1a}
\end{figure}

From the exemplary assumption about a zero-sum game, we claim that the individual payoffs will be as follows:\\
payoff(A-C)=(1-0)+(0-1)+(1-1)+(1-0)+(1-0)+(0-0)= 2,\\
payoff(A-D)=(1-1)+(0-0)+(1-1)+(1-0)+(1-1)+(0-1)= 0,\\
payoff(A-E)=(1-1)+(0-1)+(1-0)+(1-1)+(1-1)+(0-1)=-1,\\
payoff(B-C)=(0-0)+(1-1)+(1-1)+(0-0)+(1-0)+(1-0)= 2,\\
payoff(B-D)=(0-1)+(1-0)+(1-1)+(0-0)+(1-1)+(1-1)=-1,\\
payoff(B-C)=(0-1)+(1-1)+(1-0)+(0-1)+(1-1)+(1-1)=-2.\\
Real payoff data are shown in Fig 2.

\begin{figure}[h!]
\centering
{\includegraphics[width=10cm]{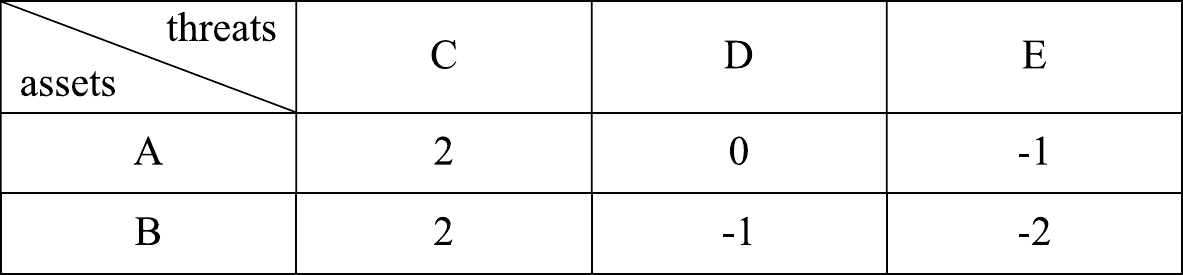}}
\caption{2x3 zero-sum game – notation with real data}
\label{fig:2a}
\end{figure}

Parameters and payoffs are usually set with greater precision, for which both total and real codes are suitable. They can be obtained on the basis of market research and marketing situation \cite{Kumar:mib}. Instead of binary logic we can use multi-valued logic, for example three-valued logic \{1,0,1\} or Lukasiewicz's logic \{0,1,2\}. 
Another approach is to use interval and fuzzy calculus \cite{chalco:mib}. For example, subtraction (used in the above formula for estimation of with drawls) on  interval basis is carried out on interval variables as follows \cite{Mandal:min}:
\begin{equation}
[x] - [y] = [\mathop x\limits_ -   - \mathop y\limits^ -  ,\mathop x\limits^ -   - \mathop y\limits_ -  ],
\end{equation}
where: $[\mathop x\limits_ -  ,\mathop x\limits^ -  ]$ - the lower and upper limits of variable x (variable interval limits [x]).
In the case when we use other payoff estimators we will use equally logically interpreted (in the proposed source) actions, for example:
\[[x] + [y] = [\mathop x\limits_ -   + \mathop y\limits_ -  ,\mathop x\limits^ -   - \mathop y\limits^ -  ],\]
\[[x]*[y] = [\min \{ \mathop x\limits_ -  \mathop y\limits_ -  ,\mathop x\limits_ -  \mathop y\limits^ -  ,\mathop x\limits^ -  \mathop y\limits_ -  ,\mathop x\limits^ -  \mathop y\limits^ -  \} ,\max \{ \mathop x\limits_ -  \mathop y\limits_ -  ,\mathop x\limits_ -  \mathop y\limits^ -  ,\mathop x\limits^ -  \mathop y\limits_ -  ,\mathop x\limits^ -  \mathop y\limits^ -  \} ],\]
\[[x]/[y] = [x]*(1/[y]),\]
\[1/[y] = O\mathop {}\limits_{} \mathop {}\limits_{} if\mathop {}\limits_{} [y] = [0,0],\]
\[1/[y] = [1/\mathop y\limits^ -  ,1/\mathop y\limits_ -  ]\mathop {}\limits_{} \mathop {}\limits_{} if\mathop {}\limits_{} 0 \notin [y],\]
\[1/[y] = [1/\mathop y\limits^ -  ,\infty [\mathop {}\limits_{} \mathop {}\limits_{} if\mathop {}\limits_{} \mathop y\limits_ -   = 0\mathop {}\limits_{} and\mathop {}\limits_{} \mathop y\limits^ -   > 0,\]
\[1/[y] = ] - \infty ,1/\mathop y\limits^ -  ]\mathop {}\limits_{} \mathop {}\limits_{} if\mathop {}\limits_{} \mathop y\limits_ -   < 0\mathop {}\limits_{} and\mathop {}\limits_{} \mathop y\limits^ -   = 0,\]
\begin{equation}
1/[y] = ] - \infty ,\infty [\mathop {}\limits_{} \mathop {}\limits_{} if\mathop {}\limits_{} \mathop y\limits_ -   < 0\mathop {}\limits_{} and\mathop {}\limits_{} \mathop y\limits^ -   > 0,
\end{equation}
where: $O$  - stands for a zero set.

The same applies when we use fuzzy variables for estimating payouts which in one of the possible variants we can treat even as a set of - sections, i.e. intervals.\\
Probability calculus can be used in various ways to determine the level of payouts. We can use Pawlak's methodology \cite{Pawlak2:mib}, \cite{Pawlak:mib}, Demster-Shafer's methodology \cite{Beynon:mib}, Bayes's determinants \cite{Josang:mib}, entropy estimation \cite{Gray:mib}, stochastic structures, etc. The choice depends on the situation and adequacy conditions and clarity of their description. 
\section{Probabilistic estimation of payoffs in relation to assets and threats}
On the basis of market research and forecasting analyses, we determine the probability of threats at any time. It is a set of probabilities which in total do not give unity because each of them is completed to unity only for the situation when the threat does not occur $pp\left( X \right) + pp\left( X \right) = 1;$  $X \in \left\{ {C,D,E} \right\}$ and thus \\
$\sum\limits_{X \in \left\{ {C,D,E} \right\}} {pp\left( X \right)}  \ne 1\ $ - Exemplary data for the market launch of VR glasses and those related to competition threats are as follows (Fig. 3):\\
\begin{figure}[h!]
\centering
{\includegraphics[width=12cm]{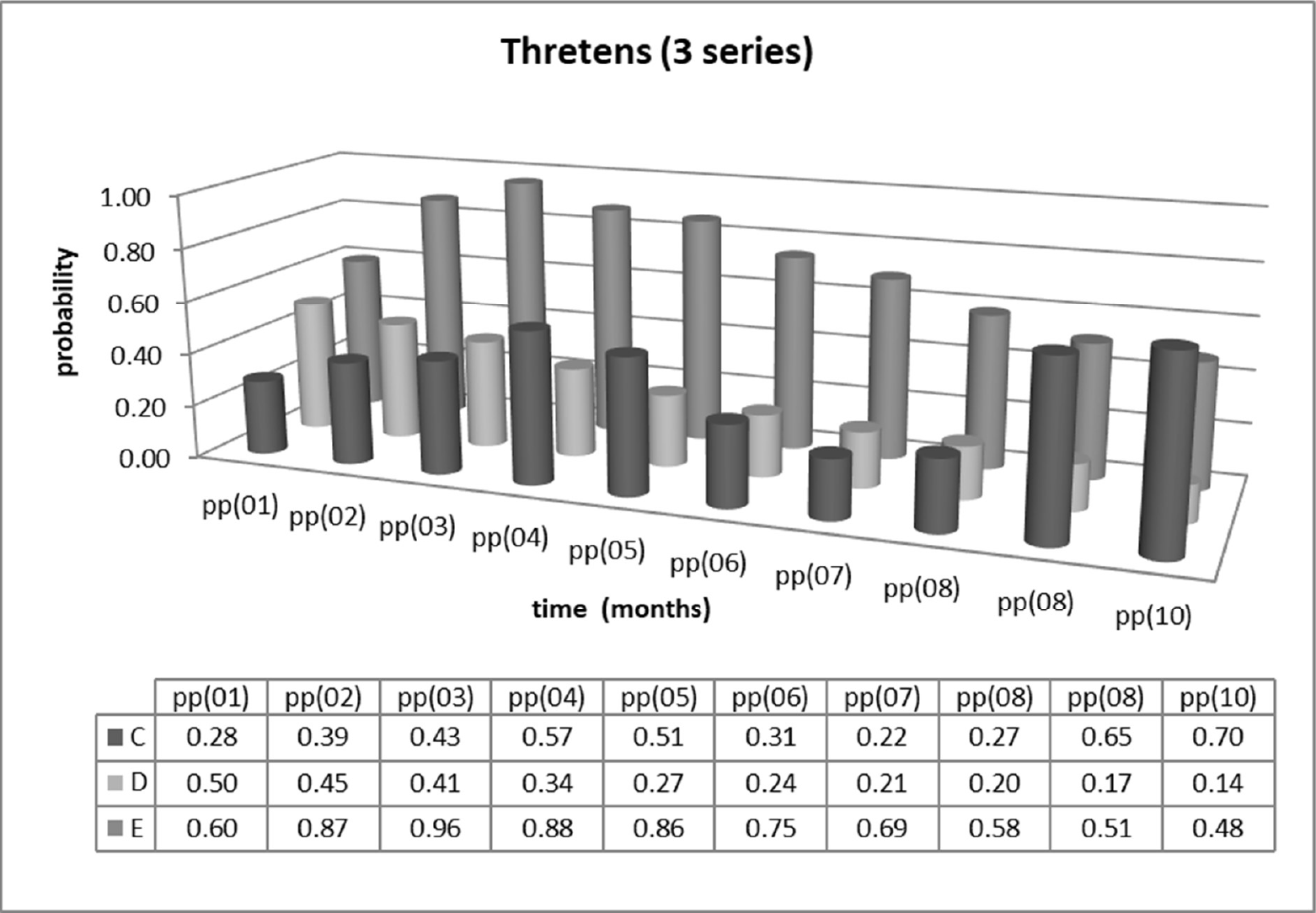}}
\caption{Distribution of probability for threats \{C, D, E\} within 10 months}
\label{fig:3b}
\end{figure}

After the normalization, we obtain the relative values of probabilities of occurrence of threats related to strategies C, D and E for 1 and 10 months (Fig. 4). Figures 3 or 4 can be used to assess payouts for strategic games every month. Payouts relating to assets should be estimated similarly. A created game may have a zero sum (as shown in the example in Fig. 1.) or not if having distributed the payouts from threats and assets \cite{Myerson:mib}, \cite{Goeree:mib}. The value of a game is valued due to monthly playing. If it is positive then we gain and the situation is beneficial to us, making the process of marketing a product profitable, if the value of the game is negative then we should change the strategy or withdraw from a project. If the data from Fig. 1 (or 2) are inaccessible to us we must estimate them ourselves and return to the parameters of strategy development (p1,...., pn) and their probabilistic features, i.e. for example the probability of competition impact, effectiveness of marketing, level of marketability of goods, possibilities of cost reduction, etc. The paper \cite{Geiler:mib} presents the method of estimating payouts based on the evaluation of benefits in the strategy with the help of the entropy formula. In such an algorithm we use a standardized structure of probabilities corresponding to parameters p1,.... pn.

\begin{figure}[h!]
\centering
{\includegraphics[width=12cm]{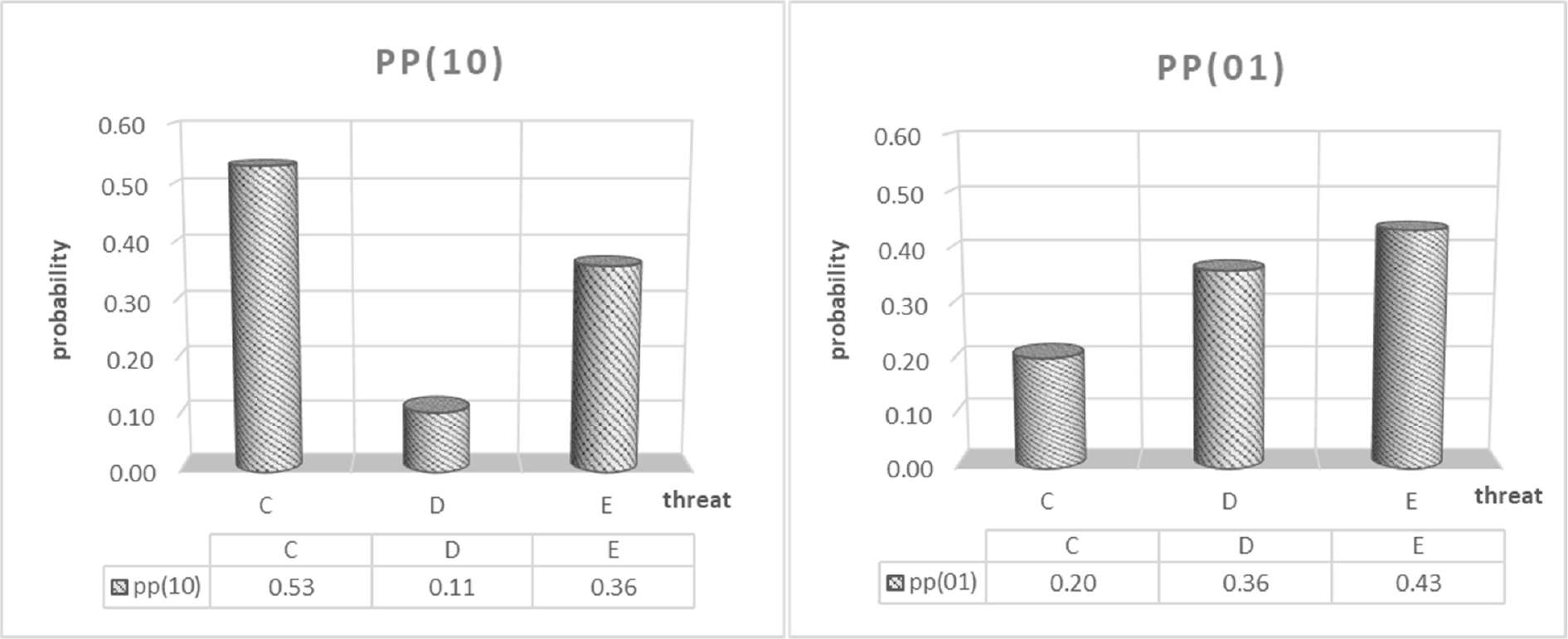}}
\caption{Effect of standardization of threat probabilities for the first and last month after the product has been placed on the market}
\label{fig:4b}
\end{figure}
In the game presented in the first chapter (Fig. 2), row A dominates over row B, and column E dominates over columns C and D. Solution of the game (game value) appears at the cross-roads of row A and column E, and it is negative. This indicates that it is not economically viable to place the product on the market (or another implementation described in the same way). However, let us note that our strategic analysis is not accurate; it is based on a binary range. What will happen when we make our description more precise using a deterministic notation with the help of numbers, e.g. real numbers. Appropriate strategies will be as follows:\\
A=\{0.88; 0.24; 0.52; 0.91; 0.71; 0.02\}, instead of A=\{1, 0, 1, 1, 1, 0\},\\
B=\{0.32; 0.68; 0.53; 0.14; 0.06; 0.77\}, instead of B=\{0, 1, 1, 0, 0, 1\},\\
C=\{0.05; 0.61; 0.53; 0.12; 0.08; 0.30\}, instead of C=\{0, 1, 1, 0, 0, 0\},\\
D=\{0.81; 0.11; 0.50; 0.22; 0.72; 0.84\}, instead of D=\{1, 0, 1, 0, 1, 1\},\\
E=\{0.67; 0.72; 0.07; 0.55; 0.60; 0.53\}, instead of E=\{1, 1, 0, 1, 1, 1\}.\\  
Therefore, we receive the payoff structure as shown in Fig. 5. Now also row A dominates over row B. But column D dominates over columns C and E, giving a small “plus” (A-D = 0.08). This indicates the profitability of the project (but weak). The saddle point therefore moves as shown in Fig. 3, so we wonder whether it is worthwhile to market the product. Using the formula for the profit entropy ($\sum { - prob/{\mathop{\rm cost}\nolimits} \;} {\log _2}\left( {prob} \right);\ $ \cite{Neyman:mib}), we get the structure of the game presented in Fig. 7.

\begin{figure}[h!]
\centering
{\includegraphics[width=10cm]{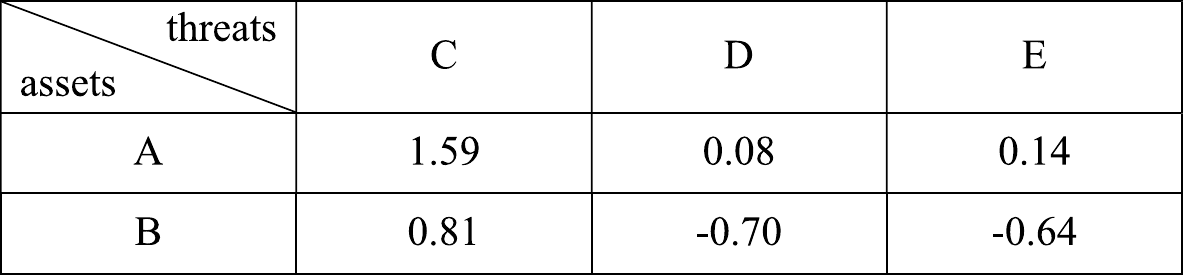}}
\caption{2x3 zero-sum game - notation with real data (on the principle of the use of parametric data in real numbers)}
\label{fig:5a}
\end{figure}

The structure based on the entropy of profit indicates the dominance of row B over A on a scale of one percent and column E over columns C and D on a scale not exceeding seven percents. The game solution is located at the crossing of strategy B and E with a negative result of -0.16. Finally, you can remove column C which in all cases was dominated, and play the game for strategy \{A, B\}\{D, E\} (Fig. 8). If there were no dominated elements, we analyze mixed strategies according to the methodology presented in \cite{Straffin:mib}:

\begin{figure}[h!]
\centering
{\includegraphics[width=8cm]{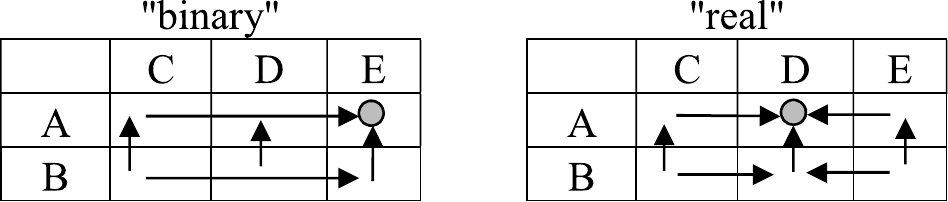}}
\caption{Saddle point movement comparison of binary and real principle - Fig. 2 and 5)}
\label{fig:6a}
\end{figure}
Let us add a new strategy X obtained for example by correcting the threats assessed using the P6 parameters in strategies C and D (C=\{0.05; 0.61; 0.53; 0.12; 0.08; 0.40\}, D=\{0.81; 0.11; 0.50; 0.22; 0.72; 0.68\}) (Fig. 9). Row B and column C are again excluded as dominated. Other elements of the game structure are not dominated, so we can use mixed strategies (Fig. 10).

\begin{figure}[h!]
\centering
{\includegraphics[width=10cm]{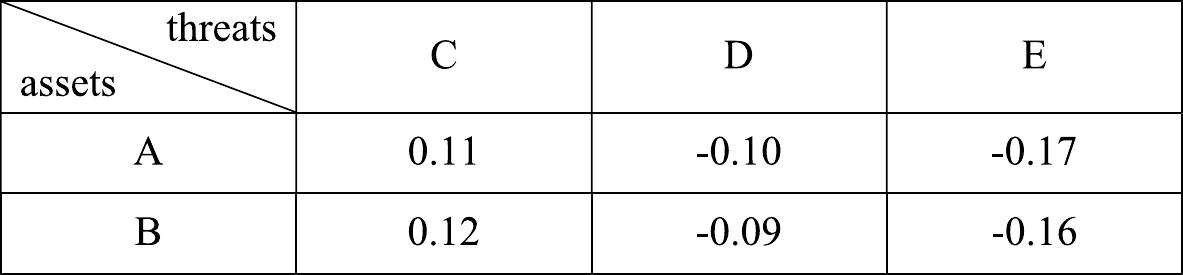}}
\caption{2x3 zero-sum game - notation with real data (on the principle of the use of profit entropy)}
\label{fig:7a}
\end{figure}
\begin{figure}[h!]
\centering
{\includegraphics[width=8cm]{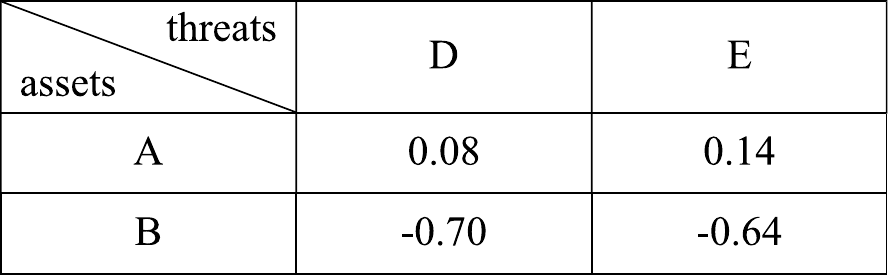}}
\caption{Reduction of the game structure by the dominated strategy C – value of the game is 0.08}
\label{fig:8a}
\end{figure}
Game value:\\
if the column plays D:   prob(A)*0.08+prob(X)*0.24=0.14\\ 
if the column plays E:   prob(A)*0.14+prob(X)*0.14=0.14

In order to obtain the positive value of the game, we have introduced an additional strategy X. This could be done by changing the values of strategic parameters p1,....,pn. Obtaining an additional strategy is a form of combating threats and increasing the effectiveness of our project. It should be remembered, however, that adjustment of strategic parameters intended for marketing, for example, is an additional cost that we will certainly incur. 
\section{Interval estimation of payoffs in relation to assets and threats}
In some situations, it may be useful to perform  interval analysis to examine the impact of changes in strategic parameters on withdrawals, and especially on the value of a game. Let's get straight to interval payouts and look for effective withdrawal  solutions in predefined intervals.

\begin{figure}[h!]
\centering
{\includegraphics[width=10cm]{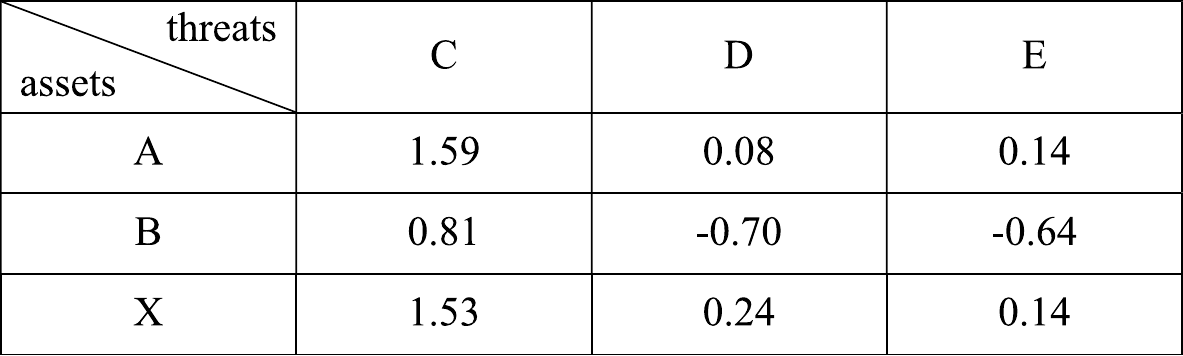}}
\caption{2x3 zero-sum game with new strategy X}
\label{fig:9a}
\end{figure}
\begin{figure}[h!]
\centering
{\includegraphics[width=8cm]{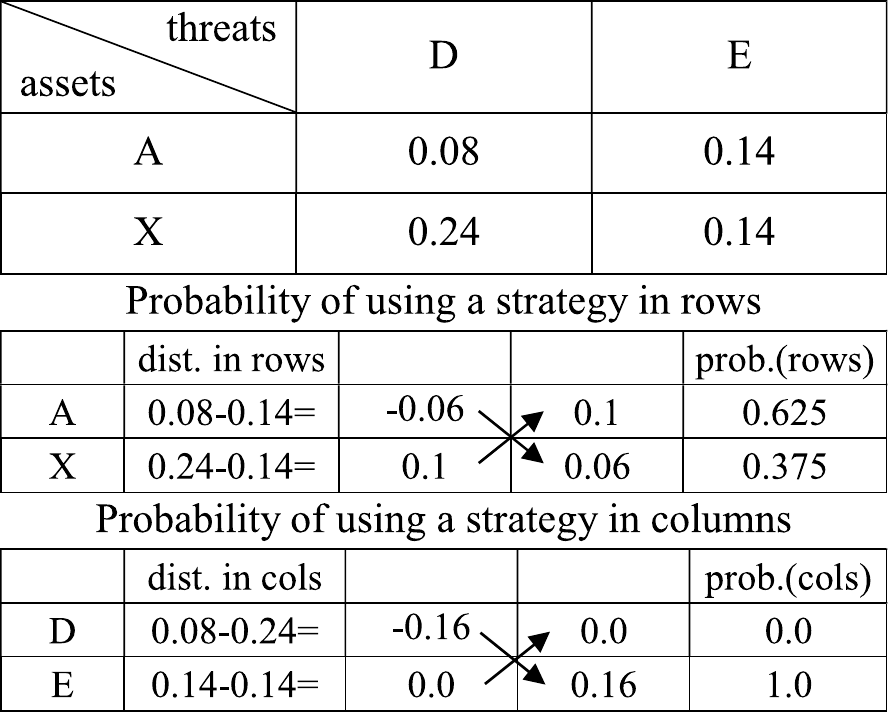}}
\caption{Use of mixedUse of mixed strategies after adding the X row option. Solution of the game with a mixed strategy}
\end{figure}
\begin{figure}[h!]
\centering
{\includegraphics[width=10cm]{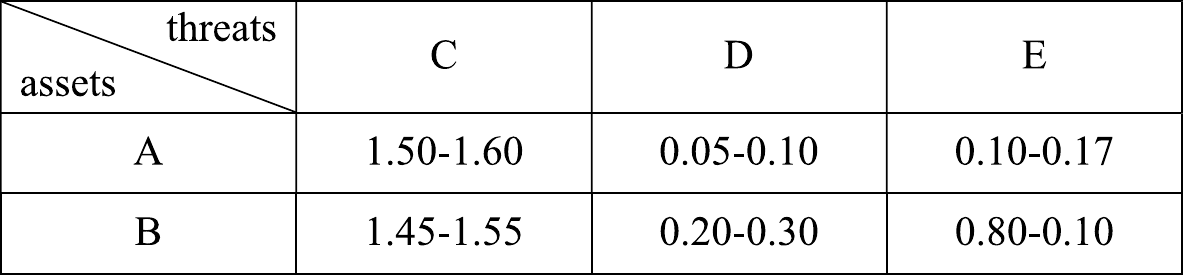}}
\caption{Interval payoff game}
\label{fig:11a}
\end{figure}
By rejecting the dominated C column and using Solver to search for better solutions, we get r the following payout structure (Fig. 12).
\begin{figure}[h!]
\centering
{\includegraphics[width=8cm]{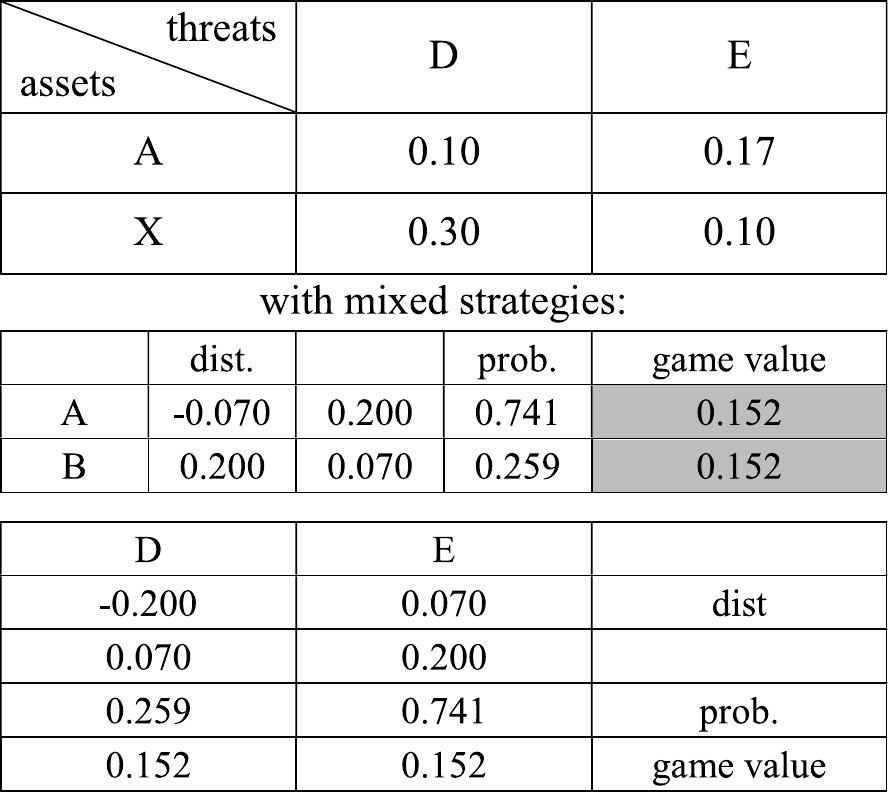}}
\caption{Searching for optimal solutions in predefined intervals}
\label{fig:12a}
\end{figure}
By changing the boundaries of ranges we can get tips on how to develop better strategies. For this to be reflected in reality, we need to introduce these strategies into a real market situation, for example, adjust  costs, change locations, change the  forms and effectiveness of the market, search for partners, etc. For example, by reducing costs we can relatively increase the A-E payout from 0.17 to 0.2 which leads to the solution of the game presented in Figure 13. In this case of cost parameter the conclusion is trivial, but playing the game can determine the scale and trends of a threat (in chronological terms). 
\begin{figure}[h!]
\centering
{\includegraphics[width=8cm]{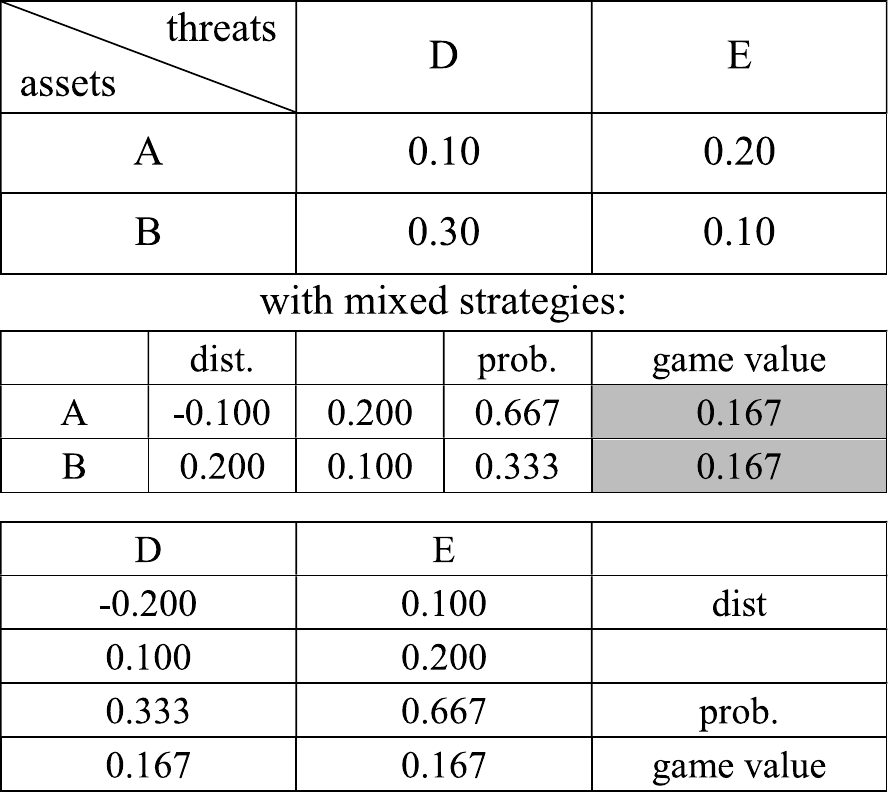}}
\caption{Effect of the impact of cost changes on the game result ($\Delta$=0.167-0.152=0.015)}
\label{fig:13a}
\end{figure}
Thanks to a well described situation relating to our market position, through proper definition and assessment of payoffs we can plan and forecast the development of investments, implementations, cooperation, cost reduction and marketing expenditures. Chronological research of the game value with threats by using and modifying our own assets gives a chance to follow changes and trends in the economic situation and to respond quickly to them. Thanks to a strategic game, which is played systematically over time, we can see the predicted threats, and also, we can examine the effects of sudden threats.

\section{Conclusion} 
The proposed players' layout is accordingly: assets – threats in a strategic game used for analysis of the market situation related to implementations, introduction of goods to the market, investments, etc. It is therefore a game of “us vs. the market situation”. The game should be played in time; it should be appropriate to corrections of our assets and emerging threats. Only the zero-sum games have been described in the simplest possible way. The payoffs were estimated on a deterministic and interval scale using probabilistically estimated parameters. Therefore, a principle was proposed and also its application manner for strategic market analysis as well as possibilities to increase the efficiency of planned projects. The proposed principle development may include non-zero sum games. 


\end{document}